\documentclass[superscriptaddress,showpacs,twocolumn,nofootinbib]{revtex4-2}

\usepackage[normalem]{ulem}
\usepackage{graphicx}
\usepackage{color}
\usepackage{verbatim}
\usepackage{subfigure}
\usepackage{amssymb}
\usepackage{amsmath}
\usepackage{comment}
\usepackage{chngcntr}
\usepackage{appendix}
\usepackage{lipsum}
\usepackage{physics}
\usepackage{mathtools}
\usepackage{xcolor}
\definecolor{LFW2024RED}{HTML}{DD4132}
\definecolor{LFW2024BLUE}{HTML}{1F5DA0}

\usepackage[
colorlinks=true,
citecolor=LFW2024BLUE, %MidnightBlue
linkcolor=MAROON, %BrickRed,
urlcolor=black,
breaklinks=true,
]{hyperref}  
\usepackage[
usenames,
dvipsnames
]{xcolor}               % additional colors

\usepackage{amsthm}
\usepackage{bm}
\usepackage[hang,flushmargin]{footmisc}
\usepackage{soul}

\usepackage{amssymb} %maths
\usepackage{amsmath} %maths
\usepackage{bm}
\usepackage{mathtools}
\usepackage{xcolor}
\usepackage{stmaryrd} %llbracket, rrbracket

\definecolor{Pantone2012}{HTML}{DD4124}
\definecolor{MAROON}{HTML}{A52A2A}

\newcommand{\si}[1]{\textcolor{black}{#1}}
\newcommand{\te}[1]{\mbox{$\mathbf{ #1 }$}}

\def\eq{\ = \ }

\def\bnabla{\boldsymbol{\nabla}}

\def\bn{\te{n}}

\def\b1{\te{1}}

\makeatletter
\newcommand{\stepsubequation}[1][]{%
  \ifmeasuring@
  \else
    \refstepcounter{parentequation}%
    \protected@xdef\theparentequation{\arabic{parentequation}}%
    \ifdefined\theHparentequation
      \protected@xdef\theHparentequation{\arabic{parentequation}}%
    \fi
    \setcounter{equation}{0}%
    \if\relax\detokenize{#1}\relax\else
      \edef\@currentlabel{\theparentequation}%
      \ltx@label{#1}%
    \fi
  \fi
}
\makeatother

\begin{document}

\title{Interfacial mass transfer resistance at fluid-fluid interfaces}
\author{Hyeongjoo Row}
\thanks{Equal contribution}
\affiliation{University of California, Berkeley, California 94720, USA}

\author{Brandon J. Wallace}
\thanks{Equal contribution}
\affiliation{University of California, Berkeley, California 94720, USA}
\affiliation{Lawrence Berkeley National Laboratory, Berkeley, California 94720, USA}

\author{${\mbox{Joshua B. Fernandes}}$}
\affiliation{University of California, Berkeley, California 94720, USA}
\affiliation{Lawrence Berkeley National Laboratory, Berkeley, California 94720, USA}

\author{Kevin R. Wilson}
\thanks{Correspondence: krwilson@lbl.gov, kranthi@berkeley.edu}
\affiliation{Lawrence Berkeley National Laboratory, Berkeley, California 94720, USA}

\author{Kranthi K. Mandadapu}
\thanks{Correspondence: krwilson@lbl.gov, kranthi@berkeley.edu}
\affiliation{University of California, Berkeley, California 94720, USA}
\affiliation{Lawrence Berkeley National Laboratory, Berkeley, California 94720, USA}

\begin{abstract}
Complex chemistry in nano- and microscale compartments is often governed by how quickly reagents transit a fluid-fluid interface. Mass transport across interfaces is commonly modeled by assuming local equilibrium, enforcing continuity of chemical potential across the interface. While adequate at large scales, this approximation may break down at the microscale, where interfacial processes can become rate-limiting.
Here, we extend linear irreversible thermodynamics to describe nonequilibrium interfacial mass transport. We identify an interface-limited regime, in which transport is governed by interfacial resistance and exhibits exponential relaxation.
Combining microfluidic and spectroscopic techniques, we introduce an experimental technique that explores this regime and provides a direct measurement of the interfacial mass transfer coefficient. For a model system consisting of acetonitrile transport across
a surfactant-stabilized water-oil interface, we obtain an interfacial transport coefficient ${M \sim 7\,{\rm nm/s}}$.
These results establish interfacial mass transfer resistance as a governing mechanism in microscale transport and provide a framework to predict, control and measure mass transport in multiphase systems at microscale.
\end{abstract}

\maketitle

\vspace{0.1in}

\noindent\textbf{\textit{Introduction.}} The transport of molecules across the boundaries of nano- and microscale objects can exert a large influence on the overall chemical evolution of a system. Mass transfer governs key steps in aerosol and cloud droplet growth~\cite{lbadaoui2021molecular}, the formation and selective partitioning of species in coacervates~\cite{zhang2024exchange} and in the chemical and material synthesis in emulsions, drug delivery and food products. 
Standard models for transport phenomena across fluid-fluid interfaces routinely assume local equilibrium at the interface. This interfacial equilibrium hypothesis implies the continuity of the thermodynamic driving variables. For example, no-slip, equal temperature, and equal chemical potential across the interface are commonly assumed for momentum, heat, and mass transfer, respectively~\cite{bird2006transport}.
However, such idealized local equilibrium assumptions may break down when interfacial transport itself is rate limiting. For momentum and heat transfer, such non-equilibrium effects have been well studied under the frameworks of the slip velocity for fluid-solid interfaces~\cite{navier1823slip} and Kapitza thermal resistance for fluid-solid~\cite{kapitza1941resistance} and solid-solid interfaces~\cite{little1959transport, swartz1989thermal}. These interfacial phenomena often become significant in small-scale systems due to the large surface-to-volume ratio. However, the role of an analogous \emph{interfacial mass transfer resistance} in multiphase mass transport still remains poorly understood, despite its potential significance in microscale mass transport~\cite{zhang2024exchange}.

In this article, we extend the framework of linear irreversible thermodynamics to consider nonequilibrium interfacial mass transport with a finite interfacial resistance. We show that such interfacial effects are characterized by a nondimensional parameter, an interfacial Biot number, which represents the relative resistances of bulk to interfacial mass transport. We also present a robust method to quantify the interfacial resistance by using a microfluidic platform that allows us to probe interface-dominated dynamics. In addition to a direct measurement of the interfacial transport resistance, our results further support recent evidence that surfactant-stabilized microfluidic droplets are not leak-proof as often assumed \cite{gruner2016controlling,etienne2018cross}. While leakage is often ignored and poorly understood mechanistically in these systems, our results indicate there is slow molecular leakage that is dominated by an interfacial mass transfer process.

\begin{figure}[h!]
\centering
\includegraphics[width=\linewidth]{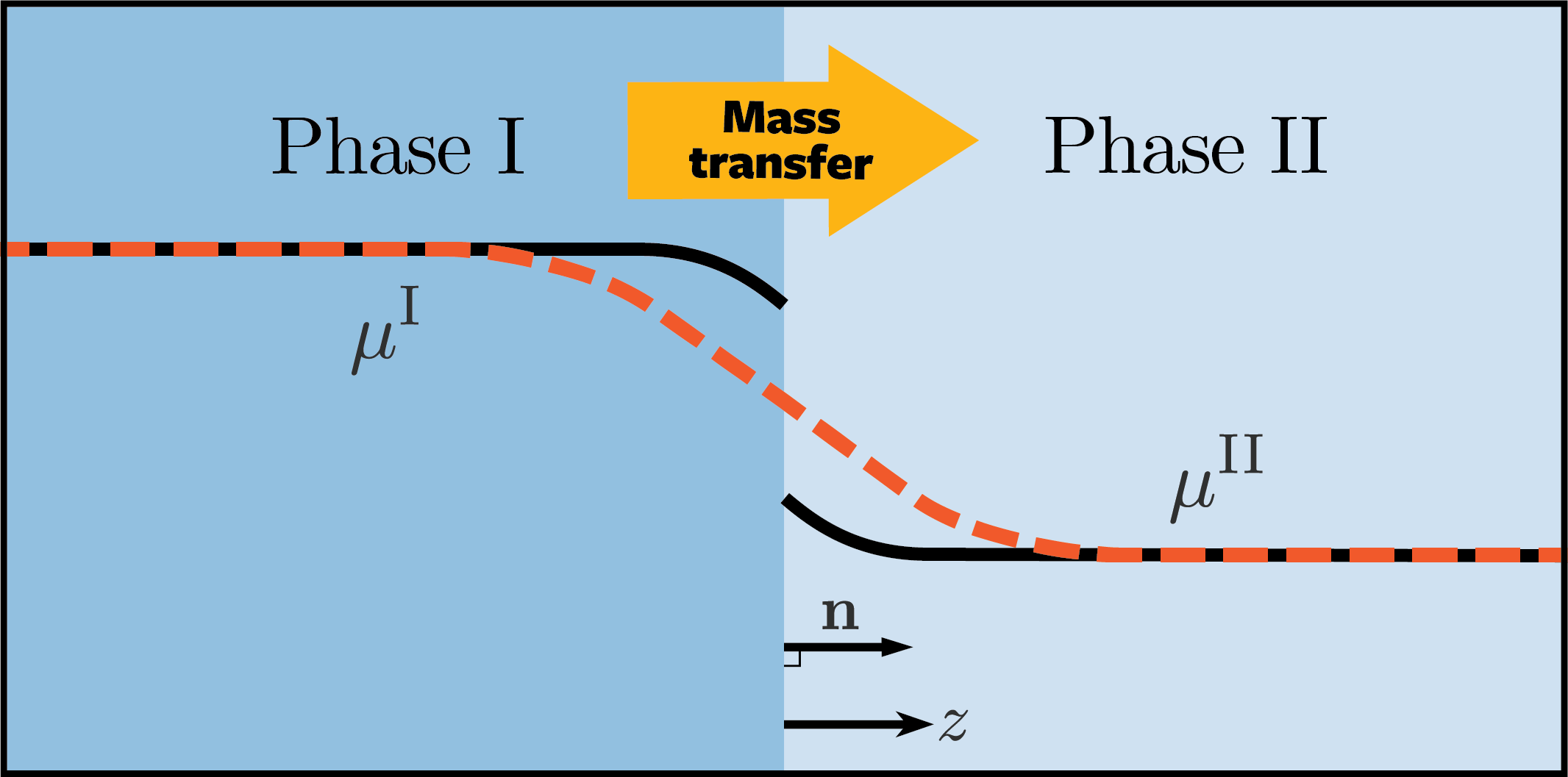}
\caption{Schematic of the interfacial mass-transfer model. Transport of a solute occurs from phase $\rm I$ (${z<0}$) to phase $\rm II$ (${z>0}$). In the classical solution-diffusion model (dashed), the chemical potential $\mu$ remains continuous across the interface (${z=0}$) and local chemical equilibrium is maintained at the interface. In contrast, introducing a finite interfacial mass-transfer resistance (solid) allows for a jump in the chemical potential.}
\label{fig1_schematic_descriptions} 
\end{figure}

\vspace{8pt}
\noindent\textbf{\textit{Theory.}} We consider mass transport across two phases, $\rm I$ (${z<0}$) and $\rm II$ (${z>0}$), as illustrated in Fig.~\ref{fig1_schematic_descriptions}. 
Within each bulk phase, mass transport is described by mass balance equations, which we close with constitutive relations for the fluxes within the framework of linear irreversible thermodynamics~\cite{prigogine_irrevthermo, degroot_noneqthermo, fong2020transport}:
\begin{align}
    \frac{\partial C_{i}^{\rm X}}{\partial t} &\eq - \bnabla\cdot \mathbf{j}_{i}^{\rm X}\ ,
    \label{eq:bulk_mass_bal}
    \\
    \mathbf{j}_{i}^{\rm X} &\eq - \sum_{j} L_{ij}^{\rm X} \bnabla \mu_{j}^{\rm X} \ ,
    \label{eq:bulk_mass_flux}
\end{align}
where $C_{i}^{\rm X}$ is the concentration, $\mathbf{j}_{i}^{\rm X}$ is the mass flux, $L_{ij}^{\rm X}$ are the Onsager transport coefficients, and $\mu_{i}^{\rm X}$ is the chemical potential. We use subscripts to index chemical species and superscripts to denote the phase. In the dilute limit (${C_{i}^{\rm X} \rightarrow 0}$), the transport coefficients reduce to ${L_{ii}^{\rm X} \approx D_{i}^{\rm X}C_{i}^{\rm X} /k_{\rm B}T }$~\cite{fong2020transport} and Eqs.~\eqref{eq:bulk_mass_bal} and~\eqref{eq:bulk_mass_flux} recover the familiar diffusion equations, where $D_{i}^{\rm X}$ is the diffusivity.

At the interface (${z=0}$), we now allow for nonequilibrium transport processes, in which the chemical potential may be discontinuous. 
Using the framework of irreversible thermodynamics~\cite{degroot_noneqthermo, fong2020transport,alkadri2025irreversible} and the modified integral theorems for surfaces of discontinuity~\cite{alkadri2025irreversible}, 
the rate of internal entropy production per unit area $\sigma^{\rm int}$ at the interface can be written as (see \si{Sec.~II.2 in Supplemental Material (SM)}~\cite{smfootnote})
\begin{equation}
\sigma^{\rm int}
\eq
-\frac{1}{T}
\sum_{i}  {\rm j}_{i} \, \llbracket \mu_{i} \rrbracket
\ \geq \ 0 \ .
\label{eq:interface_2nd_law}
\end{equation}
Here, $T$ is the temperature, ${\llbracket \mu_{i}\rrbracket\equiv (\mu_{i}^{\rm II} - \mu_{i}^{\rm I})|_{\rm interface}}$ is the jump in the chemical potential across the interface, ${{\rm j}_{i} \equiv \bn \cdot \mathbf{j}_{i}^{\rm X}|_{\rm interface}}$ is the mass flux through the interface, and $\bn$ is the surface normal to the phase interface pointing from phase $\rm I$ to $\rm II$. 
The inequality in Eq.~\eqref{eq:interface_2nd_law} represents the second law of thermodynamics for the interface, requiring the internal entropy production rate to be nonnegative.
Equation~\eqref{eq:interface_2nd_law} further reveals ${\rm j}_{i}$ and ${-\llbracket \mu_{i} \rrbracket} / T$ as the conjugate thermodynamic fluxes and driving forces at the interface, respectively. In the linear irreversible regime, we may then propose the following linear phenomenological relations between the interfacial thermodynamic forces and fluxes as
\begin{equation}
    {\rm j}_{i}
    \eq
    - \sum_{j} \ell_{ij} \llbracket \mu_{j}\rrbracket \ ,
    \label{eq:interface_mu_flux}
\end{equation}
where $\ell_{ij}$ are the interfacial transport coefficients. Equation~\eqref{eq:interface_mu_flux} can be interpreted as an extension of the Onsager constitutive relations in Eq.~\eqref{eq:bulk_mass_flux} for the bulk to those at the interface.
From Eqs.~\eqref{eq:interface_2nd_law} and~\eqref{eq:interface_mu_flux}, the square matrix composed of the coefficients $\ell_{ij}$ must be positive semi-definite, implying ${\ell_{ii}\geq0}$ and ${\sum_{i,j} \ell_{ij}\geq0}$. This formulation admits cross-coupling through nonzero off-diagonal coefficients~${\ell_{ij}\ ({i \neq j})}$, so that a chemical potential jump in one species can also drive the transfer of others across the interface, where ${\ell_{ij}=\ell_{ji}}$ by the Onsager reciprocal relations~\cite{onsager1931reciprocal1, onsager1931reciprocal2}.
While Eq.~\eqref{eq:interface_mu_flux} provides general constitutive relations for interfacial transport, including cross-couplings among species, in this work we focus on the dilute limit, where only the diagonal terms are retained.

In the limit of dilute concentrations and small chemical potential jumps ${\llbracket \mu_{j}\rrbracket/(k_{\rm B} T)\ll1}$, the interfacial flux in Eq.~\eqref{eq:interface_mu_flux} becomes ${{\rm j}_{i}=M_{i} (K_{i} C_{i}^{\rm I} - C_{i}^{\rm II})|_{\rm interface}}$, where $M_{i}$ is the inverse interfacial mass transfer resistance and $K_{i}$ is the equilibrium partition coefficient between the two phases\footnote{The interfacial flux may be written alternatively as ${{\rm j}_{i}=M_{i}’ ( C_{i}^{\rm I} - K_{i}’ C_{i}^{\rm II})|_{\rm interface}}$, with ${K_{i}’ = K_{i}^{-1}}$ and ${M_{i}’ = M_{i} K_{i}}$ (see \si{SM Sec.~II.2}). These two forms are equivalent and reflect different choices of reference phase for defining the interfacial barrier. In the original form, $M_{i}$ is associated with transfer from phase ${\rm II}$ to phase ${\rm I}$, whereas in the second form $M_{i}’$ corresponds to transfer from phase ${\rm I}$ to phase ${\rm II}$.}.  Although this linear constitutive form has been proposed previously~\cite{scott1951diffusion, shimbashi1965mass, taylor2019quantifying, zhang2024exchange}, the interfacial transport coefficient remains poorly quantified and largely unexplored.
Several properties of $M_{i}$ are worth noting. First, its dimension is that of a velocity. Second, $M_{i}$ is a dynamical parameter that strictly affects the dynamics of mass transport, not the final equilibrium state, which is determined solely by the thermodynamic parameter $K_{i}$ and the initial state. Despite its simplicity and potential significance in controlling microscale multiphase mass transport, the value of the interfacial resistance is generally unknown, even for simple systems such as oil-water interfaces.

A dimensional analysis in \si{SM Sec.~III.1}  shows that the proposed interfacial resistance introduces a dimensionless group ${{\rm Bi}\equiv M L/D}$, where $L$ is a characteristic system length scale and $D$ is the diffusivity. Drawing an analogy to convective mass transfer at an interface~\cite{incropera1996fundamentals}, we denote this dimensionless group as ${\rm Bi}$---an interfacial mass transfer Biot number. Physically, ${\rm Bi}$ represents the ratio of the bulk resistance ($L/D$) to the interfacial resistance ($1/M$) and quantifies the dynamical effects of nonequilibrium interfacial mass transfer.

The role of ${\rm Bi}$ in the mass relaxation dynamics is analyzed in detail in \si{SM Secs.~III.2-3} for planar and spherical interfaces, considering both finite and semi-infinite domains. When the bulk resistance dominates (${{\rm Bi} \gg 1}$), the overall mass relaxation is diffusive with a timescale of ${\tau_{\rm D} = L^2/D}$. In this regime, the interface remains close to local equilibrium and the dynamics reduce to those predicted by the conventional solution-diffusion model~\cite{bird2006transport}. On the other hand, when there is a significant amount of interfacial resistance (${{\rm Bi} \ll 1}$), the mass relaxation becomes slower with a timescale ${\tau_{\rm D}/{\rm Bi}}$ (see \si{Figs.~S3 and~S10}).

In the regime ${{\rm Bi} \ll 1}$, bulk relaxation is sufficiently fast that the concentration within each phase remains spatially uniform, allowing a lumped system analysis to yield
\begin{equation}
    V^{\rm I}\frac{d C^{\rm I}}{d t} \eq
    -V^{\rm II}\frac{d C^{\rm II}}{d t} \eq
    -A M\left(KC^{\rm I} - C^{\rm II}\right) \ ,
    \label{eq:lumped_sum}
\end{equation}
where $V^{\rm X}$ is the volume of a phase and $A$ is the interfacial area. Thus, the interface-limited dynamics follow the typical first-order kinetics, 
\begin{equation}
    C^{\rm X} 
    \eq
    \left(
        C^{\rm X}_{0} - C^{\rm X}_{\infty}
    \right)
    e^{-t/\tau} + C^{\rm X}_{\infty}
    \ , 
\end{equation}
with which the bulk concentrations relax exponentially towards their equilibrium values,
\begin{equation}
C^{\rm I}_{\infty}
\eq 
\frac{C^{\rm II}_{\infty}}{ K }
\eq
\frac{V^{\rm I} C^{\rm I}_{0} + V^{\rm II} C^{\rm II}_{0}}{V^{\rm I} + K V^{\rm II}} \ , 
\label{eq:eqval}
\end{equation}
with the characteristic timescale,
\begin{equation}
    \tau \eq \frac{V^{\rm I}}{M A}\left[K+\frac{V^{\rm I}}{V^{\rm II}}\right]^{-1} \ ,
\label{eq:timescale}
\end{equation}
leading to a simplified formula that may be used to estimate the interfacial transport coefficient $M$.

\begin{figure}[t!]
    \centering
    \includegraphics[width=\linewidth]{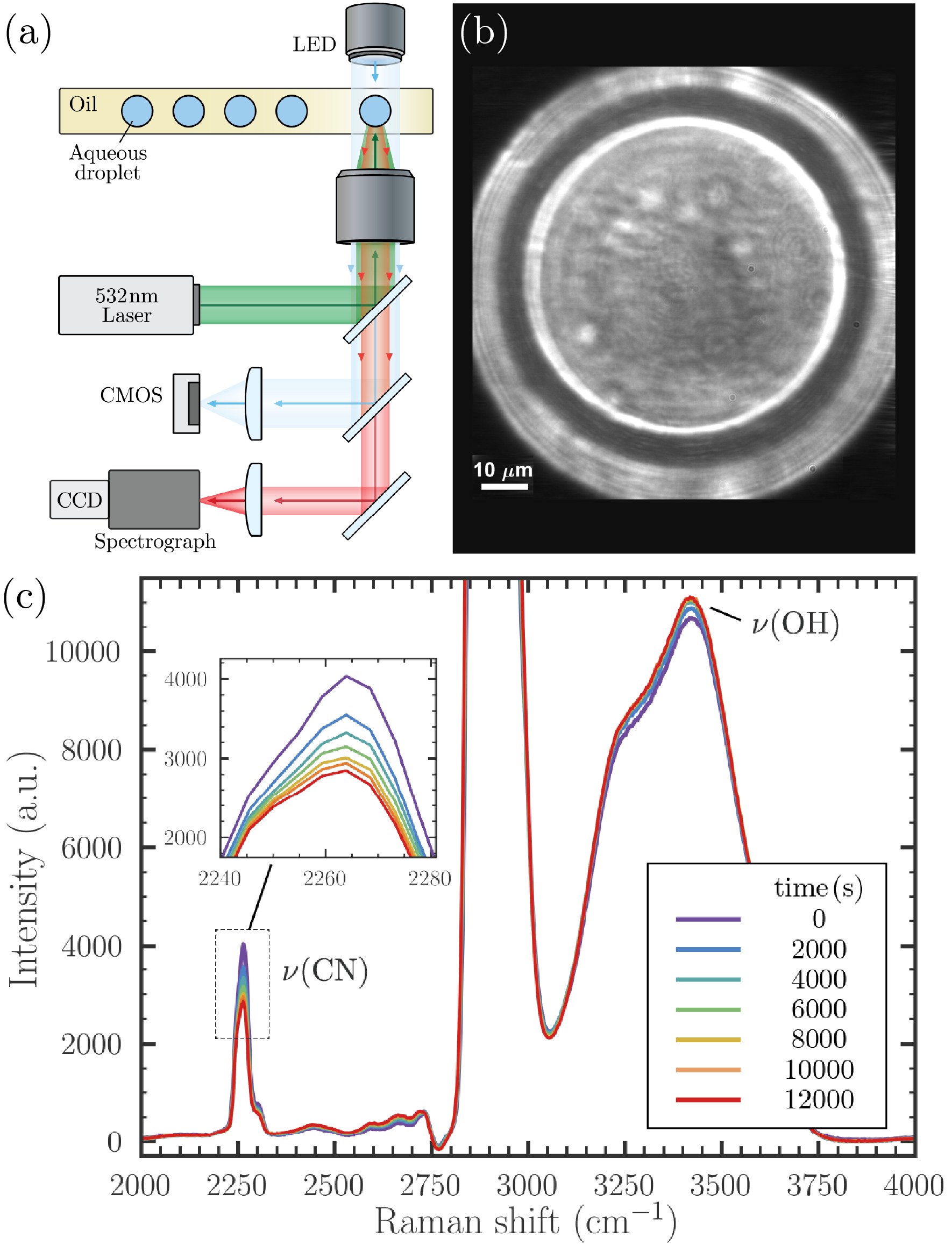}
    \caption{(a) Simplified schematic of the optical trap used to trap and isolate a single emulsion droplet and measure solute leakage \emph{in situ}. (b) Image of an optically trapped water droplet immersed in oil and surrounded by other water droplets. The RAN surfactant coats each droplet.  (c) Raman spectrum over time for droplets loaded with acetonitrile showing nitrile and water peaks at ${\nu({\rm CN}) = 2256\, {\rm cm}^{-1}}$ and ${\nu({\rm OH}) = 3450\, {\rm cm}^{-1}}$, respectively. The water peak remains stable over the experimental timescale while the nitrile peak decays, indicating solute leakage from the droplet as shown in the inset.}
    \label{fig:expt-setup}
\end{figure}
\vspace{8pt}
\noindent\textbf{\textit{Experiments.}} We now describe an experimental protocol to quantify the interfacial mass transfer resistance. As shown in the theoretical analysis, direct access to $M$ is possible only in the interface-limited regime (${\rm Bi} \ll 1$). Under these conditions, the mass-transfer dynamics exhibit simple exponential relaxation with a characteristic timescale given by Eq.~\eqref{eq:timescale}, enabling determination of $M$ from measurements of transient bulk concentrations.
Since $M$ is unknown, ${\rm Bi}$ cannot be specified \textit{a priori}. However, since ${\rm Bi}$ scales linearly with the characteristic system length, the interface-limited regime can be accessed by significantly reducing the size of the bulk domains.
Guided by this consideration, we introduce an experimental technique that employs a microfluidic platform consisting of micron-sized aqueous droplets suspended in an oil phase stabilized by a surfactant, combined with a spectroscopic technique that enables real-time measurement of solute concentrations in a droplet.

Water-in-oil emulsion droplets, with radius ${R\approx 40\, \mbox{\textmu} {\rm m}}$, are generated by using a flow-focusing junction (${50\, \mbox{\textmu} {\rm m}\times50\, \mbox{\textmu} {\rm m}}$) microfluidic chip (DropChip, Cellix Ltd.). We use a fluorinated oil (HFE7500) as the oil phase, which contains 2\% of a proprietary fluorinated surfactant (008-FluoroSurfactant, RAN Biotechnologies). 
Typical total flow rates for droplet generation are between  ${13.25\!-\!14.5\, {\mbox{\textmu} {\rm L/min}}}$, controlled by a microfluidic pump (PneuWave, CorSolutions). For analysis, the emulsions are transported to a storage microchip (Vena8 with Glass Coverslips, Cellix Ltd) with channel dimensions of $40000\, \mbox{\textmu} {\rm m} \times 800\, \mbox{\textmu} {\rm m}  \times 80\, \mbox{\textmu} {\rm m}$, which is capped to minimize potential influences from gas-phase partitioning.

To measure the leakage of an aqueous solute from an individual emulsion droplet into the surrounding oil phase, we use a modified commercial optical trap (AOT-100, Biral), which is schematically illustrated in Fig.~\ref{fig:expt-setup}(a). The detailed experimental method is described in Ref.~\cite{wallace2025enhanced} and we summarize it only briefly here. 
In short, a green (${532\,{\rm nm}}$) laser is focused through a ${100\times}$ objective to optically trap individual microemulsion droplets and isolate them from surrounding droplets within the collection microchip, with the laser power set to $106\, {\rm mW}$ for all experiments.
Backscattered Raman emissions from the trapped droplets are collected using a spectrometer and CCD camera, with an acquisition time of $2\, {\rm s}$ for all spectra.
For visualization and sizing, the droplets are illuminated using a collimated blue LED mounted above the microfluidic chip holder and droplet radii are determined from calibrated bright-field images (see Fig.~\ref{fig:expt-setup}(b)).

For leakage experiments, we choose an aqueous solution of acetonitrile (MeCN) in a surfactant-stabilized oil phase as a model system. Because MeCN has a low partition coefficient between the oil and aqueous phases, the resulting relaxation dynamics occur on experimentally accessible timescales. All experiments are performed using a $5\, {\rm M}$ bulk aqueous solution of MeCN, which is loaded into the microfluidic setup to generate emulsion droplets with an initial concentration identical to that of the bulk aqueous solution. To minimize water loss from the droplets during leakage experiments, the oil phase is saturated with water before use. After droplet generation, the temporal evolution of the MeCN concentration within each droplet is measured by Raman spectroscopy (Fig.~\ref{fig:expt-setup}(c)). We quantify the time-dependent leakage of MeCN from the aqueous microdroplet by tracking the CN stretching intensity at $2256\, {\rm cm}^{-1}$, normalized by the water band centered at $3450\, {\rm cm}^{-1}$. In the spectra shown in Fig.~\ref{fig:expt-setup}, a large peak centered at $2900\, {\rm cm}^{-1}$ is attributed mostly to the acrylic microchip and the oil phase, and is therefore excluded from the analysis. All droplet leakage experiments are repeated in triplicate.

We fit the concentration traces to an exponential decay to obtain the equilibrium signal and the characteristic relaxation time, which determine the volume ratio ${V^{\rm II}/V^{\rm I}}$ through Eq.~\eqref{eq:eqval} and the interfacial mass transfer parameter $M$ through Eq.~\eqref{eq:timescale}. For a spherical droplet of radius $R$, we use ${A=4\pi R^2}$ and ${V^{\rm I}=4\pi R^3/3}$, where phase I is the aqueous phase. We also independently measure the partition coefficient $K^{\rm OW}$ of MeCN between the oil and water using the miniature shake-flask method~\cite{cumming2017octanol}, as described in \si{SM Sec.~IV.1}. This provides all inputs needed to determine $M$ directly from the measured dynamics through Eqs.~\eqref{eq:eqval} and~\eqref{eq:timescale}.

\begin{figure}[t!]
    \centering
    \includegraphics[width=\linewidth]{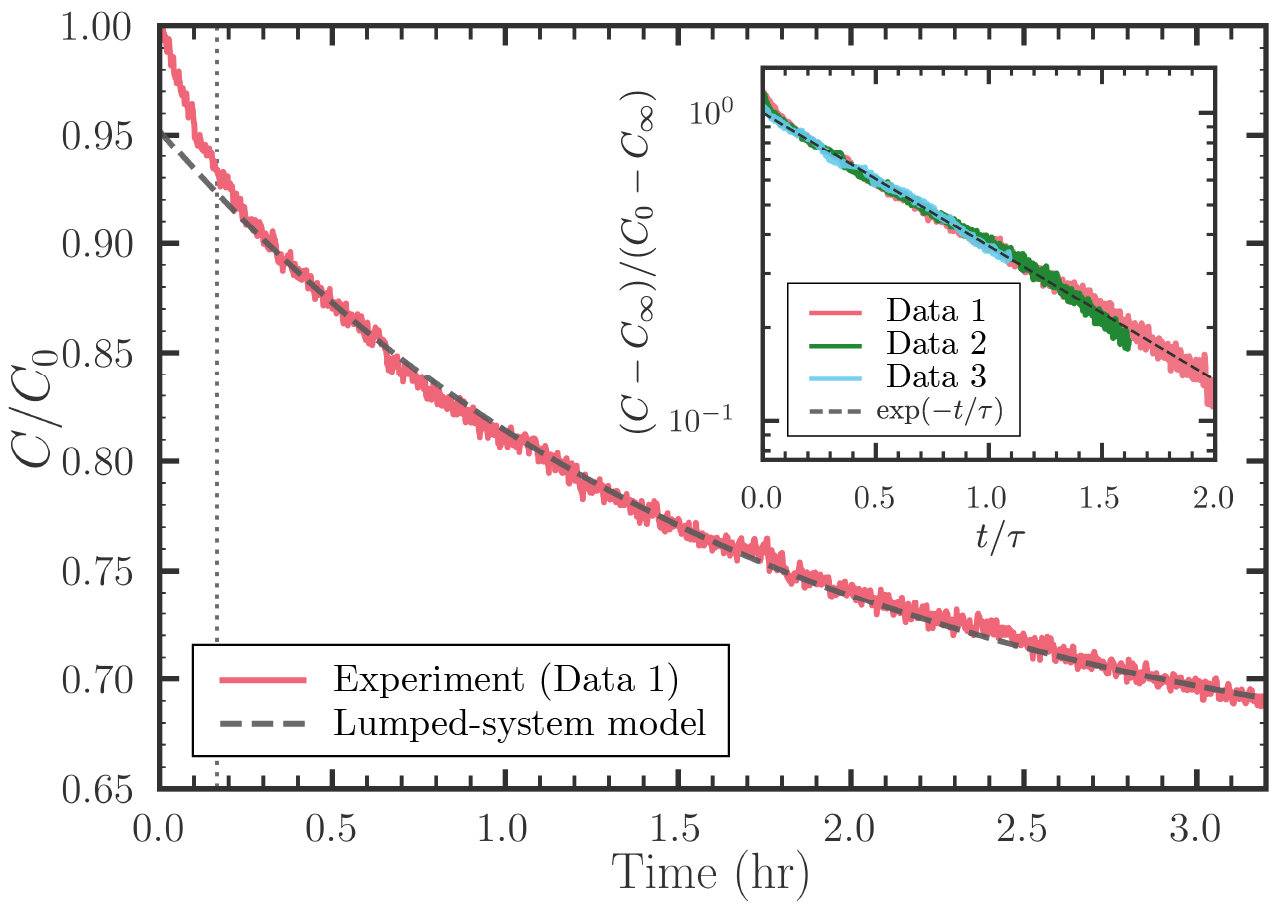}
    \caption{Experimental leakage dynamics of acetonitrile from a trapped droplet. Main panel: Data 1 (red) and an exponential fit based on the lumped system analysis (black dashed); fitting starts at the vertical gray dotted line. Inset: three independent datasets, which appear distinct in raw form, collapse onto a single master exponential curve after rescaling.}
    \label{fig:expt-results}
\end{figure}

Figure~\ref{fig:expt-results} compares the measured leakage dynamics with the fitted exponential relaxation. The fit captures the data well, except for a small deviation at the earliest times (less than ${\sim\!10}$ minutes), before the fitting window. This initial non-exponential behavior is due to the finite timescale of bulk diffusion, during which concentrations within each phase have not yet become spatially uniform. As shown in the inset, concentration traces that appear distinct in raw form (see \si{Fig.~S17}) collapse onto a single curve after shifting and scaling the data using the fitted equilibrium signal and relaxation time. This collapse demonstrates that the interface-limited exponential model captures a common transport mechanism across independent experiments.

The measured and fitted quantities for each experiment are summarized in Table~\ref{tab:parameters}. Although the droplet radius $R$, volume ratio, and relaxation timescale vary across the experiments, the inferred interfacial mass transfer coefficient remains consistent, with ${M\approx 7\,{\rm nm/s}}$ for the acetonitrile-water-RAN-HFE system. Using this value together with the reference diffusion coefficient of MeCN in water ${D = 1.64\,{\rm nm}^{2}/{\rm ns}}$~\cite{easteal1985pressure}, we find ${{\rm Bi} \approx 10^{-4}\ll1}$, justifying the lumped description for interface-limited dynamics.

\begin{table}[t!]
    \centering
    \begin{tabular}{c c c c}
        \hline
        \textbf{\ } & \textbf{Data 1} & \textbf{Data 2} & \textbf{Data 3} \\
        \hline
        \hline
        $R$ (\textmu m)  & 40.7 & 41.5 & 45.2\\
        $V^{\rm II}/V^{\rm I}$  & 4.90 & 7.71 & 13.47  \\
        $\tau$ (hr) & 1.67 & 2.15  & 3.18 \\
        $\ \,$~$M$ (nm/s)~$\ \,$  & 7.13 & 7.38 & 7.07 \\
        ${\rm Bi}$ & $\ $~${1.77\!\times\!10^{-4}}$~$\ $ & $\ $~${1.87\!\times\!10^{-4}}$~$\ $ & $\ $~${1.95\!\times\!10^{-4}}$~$\ $ \\
        \hline
    \end{tabular}
    \caption{ The measured droplet size $R$ and fitted parameters. The relaxation timescale $\tau$ and the equilibrium signal were fitted to the data in Fig.~\ref{fig:expt-results}. Using the measured partition coefficient ${K = 0.112}$ and diffusivity ${D = 1.64\,{\rm nm}^{2}/{\rm ns}}$ from Ref.~\cite{easteal1985pressure}, we estimate the volume ratio $V^{\rm II}/V^{\rm I}$ from Eq.~\eqref{eq:eqval} and determine the mass transfer coefficient $M$ from Eq.~\eqref{eq:timescale}. The Biot number was computed as ${{\rm Bi}=MR/D}$.}
    \label{tab:parameters}
\end{table}  
\vspace{8pt}
\noindent\textbf{\textit{Conclusions.}} In this work, we extend the principle of linear irreversible thermodynamics to interfacial mass transport by suggesting interfacial Onsager transport laws through Eq.~\eqref{eq:interface_mu_flux}. Instead of assuming local equilibrium at the interface, as in the solution-diffusion model, we propose a generalization in which mass flux at an interface is proportional to the chemical driving force that is the difference in the chemical potentials across the interface. In the dilute limit, this general relation reduces to a simple linear transport law driven by the departure from interfacial equilibrium.

The relative importance of interfacial and bulk mass transfer is governed by the Biot number. In the limit of large Biot number, interfacial resistance is negligible and the solution-diffusion model is recovered. In the opposite limit of small Biot number, bulk resistance becomes negligible and the relaxation dynamics are controlled by interfacial transport. In this interface-limited regime, we develop a lumped system description which results in the exponential leakage dynamics of a solute from a micron-sized water droplet suspended in oil. Application of the analysis to the microfluidic experiments yields an interfacial mass transfer coefficient of approximately ${7\,{\rm nm/s}}$. This finite resistance causes surfactant-coated water droplets to leak solute into the oil phase on a timescale of hours, showing that such droplets are not perfectly sealed by surfactants even under unfavorable partitioning into the oil phase.

More broadly, these results show that interfacial resistance can dominate bulk diffusion in microscale systems with ${{\rm Bi}<1}$. The coefficient $M$ encodes the microscopic physics of interfacial transport, including interfacial structure and the interactions among solute, solvent, and surfactant molecules, which together define the effective free-energy landscape for crossing the interface. Furthermore, beyond the dilute limit, these interactions can produce more complex transport phenomena, particularly in concentrated or multicomponent systems where cross-species coupling (${\ell_{ij}\neq0}$ for ${i\neq j}$) may emerge at the interface. Lastly, the driving force becomes electrochemical in ionic systems, introducing additional coupling through the electric potential. Resolving these outstanding questions should constitute future work.

\vspace{8pt}
\noindent\textbf{\textit{Acknowledgments.}}  This work was supported by the Condensed Phase and Interfacial Molecular Science Program (CPIMS), in the Chemical Sciences, Geosciences, and Biosciences Division of the Office of Basic Energy Sciences of the U.S. Department of Energy under Contract No.~DE-AC02-05CH11231. H.R. and K.K.M. were partially supported for performing this work by the National Science Foundation through NSF-DFG~2223407 and the Deutsche Forschungsgemeinschaft (German Research Foundation) 509322222. H.R. acknowledges support from the National Institutes of Health (U01NS136405). J.B.F. acknowledges support from the U.S. Department of Energy, Office of Science, Office of Advanced Scientific Computing Research, Department of Energy Computational Science Graduate Fellowship under Award Number DE-SC0023112. The authors are also grateful to Ashok Ajoy and Karthik Shekhar for their support throughout this work.

\bibliography{references}
% Produces the bibliography via BibTeX.

\end{document}